\documentclass{rspublic}

\usepackage{graphicx}
\usepackage{bm}
\usepackage{float}
\usepackage{subfigure}
\usepackage{amsmath}
\usepackage{amsfonts}

\begin{document}

\title[Magnetic Fields]{
Magnetic Fields in the Aftermath of Phase Transitions
}

\author[T. Vachaspati]{Tanmay Vachaspati}

\affiliation{
Institute for Advanced Study, Princeton, NJ 08540, USA\\
and\\
CERCA, Department of Physics,\\
Case Western Reserve University, 10900 Euclid Avenue,\\
Cleveland, OH 44106-7079, USA
}

\label{firstpage}
\maketitle

\begin{abstract}{Place keywords here}
The COSLAB effort has focussed on the formation of topological
defects during phase transitions. Yet there is another potentially
interesting signature of cosmological phase transitions, which 
also deserves study in the lab. This is the generation of magnetic
fields during phase transitions. In particular, cosmological
phase transitions that also lead to preferential production
of matter over antimatter (``baryogenesis''), are expected to
produce helical (left-handed) magnetic fields. The study of
analogous processes in the lab can yield important insight
into the production of helical magnetic fields, and the 
observation of such fields in the universe can be
invaluable for both particle physics and cosmology. 
\end{abstract}

\section{Introduction}

Phase transitions are a common theme in condensed matter
systems and cosmology. In the laboratory, the experimental
condensed matter physicist studies the response of materials
under changes in external parameters such as temperature,
pressure, and magnetic field, often observing sharp changes
in certain properties that signal a phase transition. In
cosmology, we do not have the luxury of a controlled
experiment. Yet it seems clear, based on the success of big
bang nucleosynthesis, that the expansion of the universe
caused the universe to cool down and that matter within the
early cosmos could also have gone through phase changes.
We are now situated in a low temperature environment and
can only deduce the occurrence of a phase transtion by
studying its aftermath.

Luckily, a wide class of phase transitions leave a tell-tale
remnant in the form of topological defects. They are ``defects''
because they indicate regions of the system that were unable
to complete the phase transition. They are ``topological''
because the reason for the incomplete phase transition is
due to the topology involved in the phase transition.
Topological defects in the universe could manifest themselves
as magnetic monopoles, cosmic strings and domain walls,
and there is an ongoing search for these structures in 
astronomical and cosmological surveys.
As we have heard at this meeting, similar searches
in condensed matter have met with success in a variety
of systems, and there is effort to quantify the production
of topological defects at a phase transition.

My talk is about phase transitions in which the requisite
topology for defects is absent. Such phase transitions
are certainly relevant in the context of cosmology because
the standard model of the electroweak interactions
has a phase transition without topological complications. 
Is there any hope of observing the aftermath of such phase 
transitions?

I shall argue that cosmic magnetic fields may be remnants from
a phase transition. Further, in certain settings, the magnetic
fields are helical, their non-trivial helicity having
its origin in the CP violation present in particle physics.
So the detection of cosmic magnetic fields and their helicity
would allow us to infer a cosmological phase transition and
provide us with an alternative probe of fundamental CP violation.
Just as for topological defects, the laboratory may be a
convenient setting to study magnetic field generation
during phase transitions. Whether this is feasible would
depend on how far we can prepare analogies between 
the cosmological and condensed matter systems.

\section{Magnetic fields from cosmic phase transition}

There are several alternative paths to deduce that magnetic
fields must be left-over after a phase transition. The
first is in analogy with the formation of topological
defects. During the phase transition, the order parameter 
must take on a value independently in every correlation
size domain. Denoting the order parameter by $\Phi$, this
implies non-zero spatial gradients, that is, 
$\nabla \Phi \ne 0$. The gradient energy density, however, 
can still vanish since it is given by $|D_\mu \Phi|^2$ 
and the gauge fields may cancel out the spatial gradients,
giving $D_\mu \Phi =0$. At a phase transition, though, the 
system is thermal and the energy density does not vanish.
On dimensional grounds we expect 
$D_\mu \Phi \sim T_{\rm pt}^2$ where $T_{\rm pt}$ 
is the temperature at which the phase transition occurs.
Consequently, we also expect there to be contribution
to the energy density coming from the electric and magnetic
field strength. As a result, we expect some magnetic
field to be left-over after the phase transition, and
since this field is embedded in a highly conducting
plasma, some of it will remain frozen-in (Vachaspati, 1991). 

The same conclusion can be reached by realizing that
even if a model does not contain topological defects,
it will almost certainly contain ``embedded defects'',
as is the case in the standard electroweak model
(Ach\'ucarro \& Vachaspati, 2000).
These are unstable solutions in the model and can be
in the form of magnetic monopoles. Such unstable objects 
can be produced by the Kibble mechanism and will decay
soon thereafter. 
However, in this process, the embedded magnetic monopoles 
will leave behind their magnetic field, which is frozen
in the ambient plasma (Vachaspati, 1994$b$).

These considerations imply a weak magnetic field after
a phase transition. Also, the coherence scale of the
magnetic field is quite small. Yet the process may still
be of relevance to the generation of galactic magnetic
fields provided we make rather optimistic assumptions 
about the galactic dynamo. 

I would now like to discuss a somewhat different approach
to the generation of primordial magnetic fields, one which 
I believe is more promising as it leads to stronger, more 
coherent fields. Also, the mechanism is richer as it shows 
that there may be a remarkable connection between baryogenesis 
at phase transitions similar to the electroweak transition 
and the helicity of cosmic magnetic fields (Cornwall, 1997;
Vachaspati, 2001). Hence, in the tradition of bringing
together particle physics and astrophysics, witness dark 
matter and dark energy, astronomical observables may 
provide yet another tool to study baryogenesis and CP 
violation in particle processes. 

\section{Elements of baryogenesis}

Particle physics, as we know it today, is almost 
completely symmetric in matter and antimatter. Thus
we would expect that the universe should be composed
of equal amounts of hydrogen and anti-hydrogen. However,
this is not the case. Galaxies are seen to collide but
never annihilate. Cosmic rays arriving to us from 
cosmological distances are mostly matter and only about
0.01\% antimatter, consistent with what is expected due
to secondary production. There are also strong 
constraints on scenarios that assume a domain structure 
for the distribution of matter and antimatter, since there
would be $\gamma -$ray production due to annihilation
at the domain boundaries. (Cohen, De R\'ujula \& Glashow, 
1998).

The overwhelming preponderance of matter in the universe
was addressed by Sakharov and can be understood if
three conditions, now known as the ``Sakharov conditions'',
are met. These are that fundamental particle physics
should contain violations of charge conjugation (C)
and charge-parity conjugation (CP), it should allow
for the conversion of antimatter to matter, and there
should be a period in cosmology where thermal equilibrium
is not maintained. Interestingly, all three of Sakharov
conditions are met in the standard model of particle
physics within standard cosmology (for a review, see
Riotto \& Trodden, 1999).

From the viewpoint of generating magnetic fields, the most 
important ingredient of the standard model is that it 
allows for transitions between matter and antimatter via 
a particular mechanism that is suppressed at low energies, 
since it proceeds by quantum tunneling, but can become 
important at high temperature (Kuzmin, Rubakov \& 
Shaposhnikov, 1985), where it is known as a ``sphaleron 
transition'' (Manton, 1983; Klinkhamer \& Manton, 1984).

Assuming the standard model of electroweak interactions,
violations of CP and departures from thermal equilibrium
at the electroweak phase transition fall short of what
is needed to explain the amount of matter-antimatter
asymmetry required for the synthesis of light elements 
(``big bang nucleosynthesis'').
The required antimatter to matter ratio is about 1
part in $10^9$ whereas the standard electroweak model 
leads to a ratio more like 1 part in $10^{20}$.
This suggests that particle physics
needs to be extended beyond the standard model. It 
is quite possible that baryogenesis in the correct
particle physics model will continue to be via sphaleron 
or sphaleron-like configurations. We will proceed under 
this assumption.

\section{More on the sphaleron}

Baryon number is classically conserved in the electroweak 
model but quantum anomalies spoil this conservation.
One way to understand this situation is that the
electroweak theory contains an infinite set of
gauge vacua, each labeled by a topological index (see
figure 1). Low energy classical
dynamics occurs within one of these topological
sectors; quantum dynamics allows for tunneling
between different vacua. In the tunneling process
from one vacuum to another, the fermions respond by
producing baryons or antibaryons. Since this is a
tunneling process, the rate of baryon number violation
is exponentially suppressed, and makes it irrelevant
for the generation of matter in cosmology.

\begin{figure}
\begin{center}
  \includegraphics[width=4.0in]{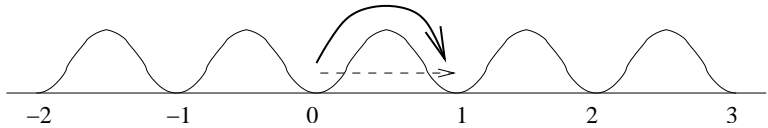}
  \caption{Periodic potential in electroweak gauge theory.
Each minima corresponds to a topological winding of the
gauge fields. Transitions from one vacuum to another
can proceed by tunneling which is very suppressed,
or over the barrier in a thermal system at high temperature. 
Such transitions
involve changing the electroweak gauge (and other) fields
through non-minimal energy. The top of the barrier 
corresponds to a solution in the electroweak model
having the minimal energy necessary to make the transition
from one vacuum to another. This solution is known as
the sphaleron. Changes in the vacuum either by tunneling
or by sphaleron transitions lead to changes in the baryon 
number.
  }
\end{center}
  \label{perpot}
\end{figure}

At high energies, the transition from one vacuum to
another may proceed without the need for tunneling,
as depicted in Figure 1. Now the path
goes over the barrier separating the two vacua. The
top of the barrier corresponds to a solution of the
field equations that has precisely one unstable 
decay mode that causes it to ``fall'' into one or
the other vacuum. This solution is called a
``sphaleron'' from the Greek roots meaning ``to fall''.
Baryons are produced/annihilated as the sphaleron decays. 

Since the early universe was very hot, sphaleron 
transitions between vacua containing different 
number of baryons are not supressed and can proceed fast 
enough to be relevant cosmologically. If there were no
CP violation, $N$ sphaleron transitions would cause 
an excess of baryons or antibaryons simply due to 
$\sqrt{N}$ fluctuations. This would imply a domain
structure where some regions of space are dominated
by baryons and others by antibaryons. Such a domain
structure is not observed and it is necessary that
CP violation be present to ensure that baryons dominate
over antibaryons throughout the universe. With CP 
violation present in the model, sphaleron transitions 
proceed more efficiently in one direction than the
other and produce more baryons than antibaryons. 

%Let us denote
%\begin{equation}
%\delta = \frac{n_b- {\bar n}_b}{n_b + {\bar n}_b}
%\end{equation}
%where $n_b$ and ${\bar n}_b$ are the number densities
%of baryons and antibaryons.

\section{Decay of the sphaleron}

We now examine the decay of a sphaleron more closely.
As we will see, this not only changes the baryon number,
but also produces helical magnetic fields. This observation
allows us to connect the observed baryon number with the 
helicity of a magnetic field.

\begin{figure}
\begin{center}
  \includegraphics[width=3.0in]{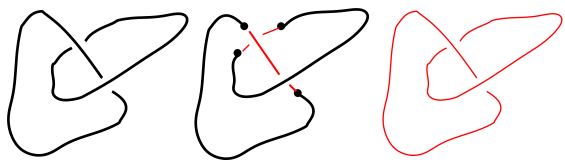}
  \caption{
This sequence of figures shows  a possible decay mechanism 
for two linked loops of electroweak Z-string (heavy curves in
first drawing). The Z-strings can break by the formation of 
magnetic monopoles and an electromagnetic magnetic field 
connects the monopole-antimonopole pairs (second drawing).  
The Z-strings can shrink and disappear, leaving behind two 
linked loops of electromagnetic magnetic field depicted by 
lighter curves in the third drawing. The initial linked loops 
of Z-string are related to the sphaleron; the final state
consists of helical magnetic fields.
  }
\end{center}
  \label{stringdecay}
\end{figure}

A heuristic way to connect the sphaleron to magnetic
fields is to use the relationship between the sphaleron
and magnetic monopoles and electroweak strings 
(Nambu, 1977; Vachaspati, 1992; Ach\'ucarro \& Vachaspati,
2000). Then the sphaleron may be thought of as two loops 
of electroweak string that are linked together, as in 
Figure 2 (Vachaspati \& Field 1994a; Hindmarsh \& James, 
1994; Garriga \& Vachaspati, 1995). 
The magnetic flux in the electroweak strings
is $Z$ magnetic flux, where $Z$ is the electroweak gauge
field, and is not related to the electromagnetic gauge
field that we shall denote by $A$. The decay of the
sphaleron now corresponds to a decay of the strings.
This can proceed in many ways but one channel is 
where the strings break by formation of (electromagnetic)
magnetic monopoles and then shrink. 
If the magnetic fields are frozen in an ambient
plasma, what remains from this process is two linked 
loops of electromagnetic magnetic field. These carry
magnetic helicity, defined as
\begin{equation}
{\cal H} = \int d^3x ~ {\bf A}\cdot {\bf B}
\end{equation}
where the integral is over all space.

The heuristic picture of the decaying sphaleron can be
confirmed by evolving the electroweak equations of 
motion with a sphaleron as initial condition. Since the
sphaleron is a static solution of the electroweak equations
of motion, it is necessary to perturb it so that it then
decays. It is quite possible that different initial 
perturbations will lead to different outcomes. However,
what is important is that the decay due to some large 
class of perturbations end up by producing helical 
magnetic fields. In Copi et al (unpublished, 2008), we 
have studied the decay of a sphaleron by numerically evolving 
the lattice electroweak equations by building on earlier work 
by Ambjorn et al. (1991), Moore (1996), Tranberg \& Smit (2003), 
Garcia-Bellido, J., Garcia-Perez, M. \& Gonzalez-Arroyo, A.
(2004), and Graham (2007). (A similar analysis has also
been done by Diaz-Gil et al (2007).) In the numerical
analysis we have implemented  absorbing boundary conditions 
by extending the scheme of Olum \& Blanco-Pillado (2000) to
non-Abelian gauge systems. The initial configuration is taken 
to be an approximate sphaleron following Klinkhamer \& Manton 
(1984).  This is not a static solution and decays. Our 
results (Figure 3) show that the Chern-Simons number decays 
as the sphaleron decays, while the magnetic helicity grows.
This numerical study confirms the heuristic picture
that sphaleron decay produces helical magnetic fields.

\begin{figure}
\begin{center}
  \includegraphics[width=3.0in,angle=-90]{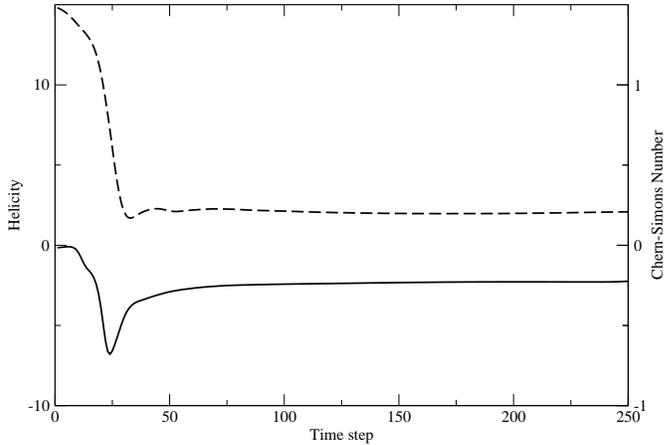}
  \caption{Magnetic helicity from the decay of a sphaleron.
The sign of the Chern-Simons number has been flipped to
make the plots clearer. 
%Energy starts leaving the simulation
%box ($128^3$) around time step of 300 and the shape of the 
%curves after this time are affected by the box boundary. 
The asymptotic helicity,
before the energy leaves the box, occurs in the time step
interval from around 50 to 250, and its value is around 
$-2.5$.  The numerical value depends on the precise decay 
channel.}
\end{center}
  \label{sphalerondecay}
\end{figure}

\section{Magnetic fields and baryogenesis}

By counting the number of baryons produced in the decay of 
the linked electroweak strings, and evaluating the helicity
of the final linked loops of magnetic field, we obtain
the relation (Cornwall, 1997; Vachaspati, 2001)
\begin{equation}
h = \frac{1}{V} \int_V d^3 x ~ {\bf A}\cdot {\bm \nabla}\times {\bf A}
  ~ \sim ~ - \frac{n_b}{\alpha}
\label{hnb}
\end{equation}

Once magnetic fields are produced, their evolution follows
the MHD equations. One well-known result is that magnetic 
helicity is conserved if the plasma has high electrical 
conductivity. The evolution
of the helical magnetic field produced by sphaleron
decay has been discussed by Vachaspati (2001). Of
particular interest is the growth of the coherence scale
of the magnetic field. Since the magnetic field carries
helicity, there is a possibility for the ``inverse
cascade'' that makes magnetic energy flow from small length 
scales to large length scales. The inverse cascade makes
the magnetic field smoother and more coherent. However,
there are also some unresolved issues in the evolution 
of the magnetic field, especially at the cosmological epoch 
when electrons and positrons annihilate and the electrical
conductivity drops quite rapidly. These are issues that are
important in the cosmological context since we are observing
the aftermath of a phase transition at very late times. In
the lab, however, these issues are not important because
we can examine the system soon after a phase transition.

\section{Detection in cosmology}

Already we have several tools to look for magnetic fields in
cosmology and we can constrain the cosmic magnetic field
strength at various coherence scales using different
observables. Big bang nucleosynthesis constrains the
magnetic field at the shortest distance scales (Kernan,
Starkman \& Vachaspati, 1996). Other measures, such as
Faraday rotation, are more sensitive to magnetic fields
with large coherence scales. Possibly there are even stronger 
constraints arising from gravitational wave production by 
non-helical magnetic fields (Caprini \& Durrer, 2002).

The future holds great promise for hunting primordial magnetic
fields. Present experiments have a snapshot of the universe
when it was 300,000 years old and what we see is the cosmic 
microwave background (CMB). The CMB propagates to us from a
distance of several Gpc ($1 ~ {\rm Gpc} \approx 10^{27} ~ {\rm cm}$).
If a cosmic magnetic field is present, the polarization of the 
CMB will undergo Faraday rotation. Ongoing experiments are
attempting to map out the CMB polarization at several different
wavelengths and we should have a detection of a cosmic magnetic
field or be able to place constraints on its strength at a 
level comparable to 1 nG.

Ultra-high energy cosmic rays (UHECRs) provide another probe of 
the cosmic magnetic field. Recently the AUGER experiment has 
identified UHECRs as protons, also attempting to correlate
the cosmic ray events with active galactic nuclei (AGNs). Since 
the UHECRs are charged, they would bend in a cosmic magnetic field
and the differences in angular position between the expected
source of the UHECR and the arrival position may be used as
a measure of the intervening magnetic field. This effort is 
still in its infancy but there is hope for the future.

While it is possible to think of many ways to detect primordial
magnetic fields, it is harder to come up with ways to detect
magnetic field helicity. Faraday rotation is only sensitive
to the line of sight component of the magnetic field but
helicity involves all components. So Faraday rotation by itself
cannot be used to detect helicity. The bending of charged particles,
as in UHECRs, is a different matter but requires proper identification
of the sources as discussed by Kahniashvili \& Vachaspati (2006).
As the cosmic ray statistics builds up, it may
eventually become possible to say something about the helicity
of the cosmic magnetic field.

\section{Elements in condensed matter systems}

The production of magnetic fields during phase transitions is
in the same general class of problems as the production of
topological defects. Yet the analogy is a little harder as
we now discuss.

In condensed matter systems, there is only one dynamical
gauge field and that is electromagnetism. This is Abelian
and not derived from non-Abelian fields above the phase
transition. So the analogy with the electroweak model
will necessarily be incomplete in this respect. On the
other hand, rather complex symmetry breakings occur in
condensed matter systems and the symmetry breaking structure
in He-3 is very similar to that in the electroweak model
(Volovik \& Vachaspati, 1996). There are also close analogs 
between (global) vortices in He-3 and Z-strings in the 
electroweak model. The production of magnetic fields in 
cosmology is similar to the production of gradients in 
one of the Nambu-Goldstone degrees of freedom. (Such 
gradients are called ``texture'' in the condensed matter 
literature.) It should be possible to experimentally check 
if textures are an aftermath of phase transition and to 
quantify them.

The production of helical texture due to quantum anomalies
may be harder to test experimentally. Yet we know that
anomalous interactions analogous to  baryon number violation
are present in He-3 (Bevan {\it et al}, 1997). Do these
interactions produce {\it helical} texture?

\section{Conclusions}

The COSLAB effort has mostly focussed on topological defects in
the aftermath of phase transitions. This has been a heroic effort,
not least because it involved bringing together cosmologists, 
condensed matter theorists and experimentalists to the same table.
It is a tribute to the COSLAB group that several difficult 
and unconventional experiments were performed. Results that 
were based on dimensional analysis and computer simulations were 
successfully tested in real systems. The ultimate theoretical 
problem of describing topological defect production at a phase
transition may require deep understanding of solitons in terms 
of particles, an unsolved problem as of now (Vachaspati, 2006).

The production of magnetic fields during phase transitions
is in the topological defects class of problems and has 
immediate application to cosmology. The puzzling aspects
of the Kibble mechanism for defect formation in gauge
systems become even more puzzling with its generalization 
to magnetic fields and would be worth testing experimentally. 
However, the cosmology and laboratory analogies for magnetic 
field production have not yet been established and will require 
further thought.

\begin{acknowledgements}
I am grateful to Ana Ach\'ucarro, Craig Copi, and 
Francesc Ferrer for collaboration and discussions.
This work was supported in part by the U.S. Department of 
Energy and NASA at Case Western Reserve University.
\end{acknowledgements}

\label{lastpage}

\end{document}